\documentclass[sigconf,nonacm,10pt]{acmart}

\usepackage{graphicx}
\usepackage{booktabs}

\usepackage{amsmath,amssymb}
\usepackage{xcolor}
\usepackage{float}

\begin{document}

\title{\textit{LIFT:} Automating Symbolic Execution Optimization with Large Language Models for AI Networks}

\author{Ruoxi Wang}
\affiliation{
  \institution{Northeastern University}
  \city{Boston}
  \state{MA}
  \country{USA}
}
\email{wang.ruoxi2@northeastern.edu}

\author{Kun Li}
\affiliation{
  \institution{Shandong University}
  \city{Jinan}
  \state{Shandong}
  \country{China}
}
\email{kunli@sdu.edu.cn}

\author{Minghui Xu*}
\affiliation{
  \institution{Shandong University}
  \city{Jinan}
  \state{Shandong}
  \country{China}
}
\email{mhxu@sdu.edu.cn}

\author{Yue Zhang*}
\affiliation{
  \institution{Shandong University}
  \city{Jinan}
  \state{Shandong}
  \country{China}
}
\email{zyueinfosec@gmail.com}

\author{Kaidi Xu}
\affiliation{
  \institution{Drexel University}
  \city{Philadelphia}
  \state{PA}
  \country{USA}
}
\email{kx46@drexel.edu}

\author{Chunchi Liu}
\affiliation{
  \institution{Huawei Technologies Company Ltd.}
  \city{Shenzhen}
  \state{Guangdong}
  \country{China}
}
\email{liuchunchi@huawei.com}

\author{Yinhao Xiao}
\affiliation{
  \institution{Guangdong University of Finance and Economics}
  \city{Guangzhou}
  \state{Guangdong}
  \country{China}
}
\email{20191081@gdufe.edu.cn}

\author{Xiuzhen Cheng}
\affiliation{
  \institution{Shandong University}
  \city{Jinan}
  \state{Shandong}
  \country{China}
}
\email{xzcheng@sdu.edu.cn}

\thanks{* Corresponding authors}

\begin{abstract}
Dynamic Symbolic Execution (DSE) is a key technique in program analysis, widely used in software testing, vulnerability discovery, and formal verification. In distributed AI systems, DSE plays a crucial role in identifying hard-to-detect bugs, especially those arising from complex network communication patterns. However, traditional approaches to symbolic execution are often hindered by scalability issues and inefficiencies, particularly in large-scale systems. This paper introduces LIFT (Large-language-model Integrated Functional-equivalent-IR Transformation), a novel framework that leverages Large Language Models (LLMs) to automate the optimization of Intermediate Representations (IRs) in symbolic execution. LIFT addresses the challenges of symbolic execution by providing a scalable, context-sensitive solution for IR transformation. The framework consists of two phases: IR Analysis and Optimization, where LLMs optimize time-intensive IR blocks, and Symbolic Execution and Validation, which includes benchmarking and semantic verification to ensure correctness and generalizability. Experiments on real-world binaries demonstrated significant performance improvements, including a 53.5\% reduction in execution time for bigtest and a 10.24\% reduction for random, along with reductions in IR statements, PUT instructions, and temporary variables. These results demonstrate that LLMs simplify IRs while maintaining functional correctness, enhancing symbolic execution in distributed AI systems.
\end{abstract}


\maketitle

\section{Introduction}
Dynamic symbolic execution (DSE) has emerged as a foundational technique in modern program analysis, now widely adopted in software testing, vulnerability discovery, and formal verification~\cite{b1,b2}. In the context of large-scale distributed systems, such as those used in AI and machine learning, DSE plays a crucial role in uncovering hard-to-detect bugs in networked systems~\cite{stoenescu2016symnet,song2014symbexnet}. By alternating concrete and symbolic execution, DSE systematically explores paths to ensure correctness in distributed AI systems, especially those with complex communication patterns.

Although there are many ways to implement DSE, a popular approach is to leverage intermediate representations (IRs) to enable platform-independent binary analysis. In distributed network environments, particularly those supporting large-scale AI training, DSE tools such as \textit{angr}~\cite{b2} and newer hybrid fuzzing frameworks like \textit{SymFusion}\cite{b3} utilize IRs to analyze binaries in a uniform and structured manner. These representations abstract low-level machine code into a normalized form, capturing essential semantics such as control flow, memory access, and arithmetic operations~\cite{b4}.

One of the central challenges in this domain is the complexity introduced by traditional compiler optimizations, which primarily focus on improving runtime performance without considering symbolic tractability~\cite{b5,b6}.
Prior work has explored symbolic-aware IR rewriting to alleviate the challenges associated with symbolic execution in modern program analysis~\cite{b8}. Most rely on manual rule-based conversion, which requires applying predefined transformations to improve symbolic reasoning. While such manual conversion techniques can yield promising results for specific cases, they face significant scalability issues. As the complexity of the target programs increases, these methods become increasingly difficult to manage, as they require extensive expert knowledge to tailor the rules for various types of code and execution paths~\cite{b7,b8}. 




Recognizing these limitations, we introduce an approach leveraging large language models (LLMs)~\cite{yao2024survey,yan2024protecting,cheng2025say,yan2025protecting,cheng2024autoiot,li2024attention}, which have shown strong capabilities in manipulating structured code~\cite{b9,b10}. LLMs automatically can produce context-sensitive IR transformations, offering a scalable solution for symbolic execution and introducing NLP into program analysis. 
Build upon this idea, we present an IR optimization framework named \textsf{LIFT} (\textbf{L}arge-language-model \textbf{I}ntegrated \textbf{F}unctional-equivalent-IR \textbf{T}ransformation). \textsf{LIFT} consists of two phases: IR Analysis and Optimization, and Symbolic Execution and Validation. In the first phase, we symbolically execute the program to extract IR blocks (IRSBs) and measure their performance using context-aware instrumentation. The most time-intensive blocks are prioritized for optimization and transformed using an automated LLM-powered pipeline. In the second phase, execution benchmarking evaluates optimization effectiveness by comparing runtime, path coverage, and solver invocation statistics, while also collecting static metrics such as IR statement count and instruction types. This step ensures that optimizations are not only effective but also generalizable for future reference. Semantic verification follows to confirm that the transformed IR remains functionally equivalent to the original, with LLMs facilitating this process for scalability and generalizability.


To demonstrate the power of our approach, we evaluated \textsf{LIFT} on real-world binaries. 
The experiments demonstrated consistent performance improvements and structural simplifications across the benchmark programs. For example, \textit{bigtest} saw a 53.5\% reduction in execution time, with 217 fewer IR statements and 106 fewer PUT operations. Random experienced a 10.24\% reduction in execution time, along with 35 fewer IR statements and 15 fewer temporary variables. Other programs like \textit{matrix} and \textit{methcall} showed more moderate improvements, with execution time reductions of 1.02\% and 1.72\%, respectively. Even \textit{bigprog} had a 0.75\% improvement in execution time, alongside 175 fewer IR statements and 85 fewer PUT instructions. Across the board, IR statements were reduced by 90 to 222, PUT instructions dropped by up to 109, and over 300 temporary variables were eliminated in certain cases. All optimizations maintained semantic correctness, confirming the effectiveness of LLMs in simplifying IRs while preserving the programs' original functionality.


Our contributions are summarized as follows:
\vspace{-0.5em}
\begin{itemize}

\item 
\textbf{Methodological Innovation}: The paper utilizes LLMs to automate the optimization of Intermediate Representations in symbolic execution, addressing the scalability and complexity issues of traditional manual methods.

\item \textbf{Framework Implementation}: We design and implement \textsf{LIFT}. \textsf{LIFT} operates in two phases: IR Analysis and Optimization, where LLMs are used to optimize time-intensive IR blocks, and Symbolic Execution and Validation, which includes benchmarking and semantic verification to ensure the correctness and generalizability of the optimizations.

\item 
\textbf{Empirical Validation}: Experiments on real-world binaries demonstrated significant performance gains, with bigtest showing a 53.5\% reduction in execution time and random a 10.24\% reduction. Across all programs, IR statements were reduced by 90 to 222, PUT instructions by up to 109, and over 300 temporary variables were eliminated, confirming the effectiveness of LLMs in optimizing symbolic execution.
\end{itemize}

\noindent\textbf{Code Availability:} The source code for LIFT is openly accessible through the Anonymous GitHub platform, which facilitates double-blind peer review by anonymizing repository details. The anonymized repository can be accessed at the following URL:
\url{https://anonymous.4open.science/r/IR-Optimization-E336}

\begin{figure*}[t]
  \centering
  \includegraphics[width=\textwidth]{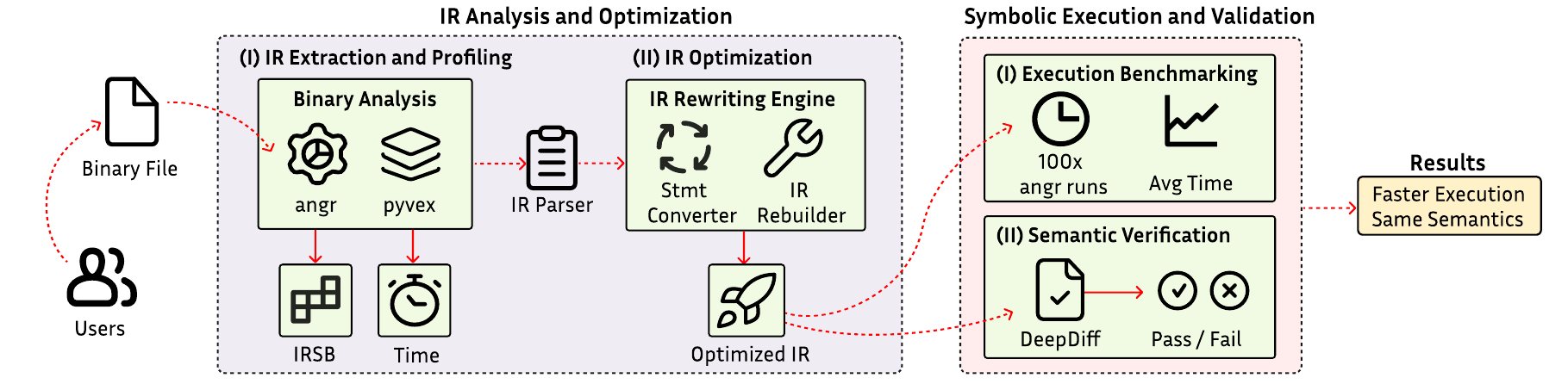}
  \Description{The architecture diagram of LIFT, showing two main stages: IR Analysis and Optimization, and Symbolic Execution and Validation, including input, LLM transformation, and benchmarking.}
  \caption{The architecture of \textsf{LIFT}}
  \label{fig:ir_manual}
\end{figure*}

\section{Background and Related Works}


\subsection{DSE for Networked Systems}

DSE is a program analysis technique used to explore execution paths, especially in complex network environments such as AI-driven distributed systems. It integrates concrete execution (running the program with actual inputs) with symbolic execution (running the program with symbolic inputs representing a range of values) to systematically explore different execution paths and uncover hidden vulnerabilities. In the context of modern AI workloads, DSE is crucial for analyzing the networked components of distributed systems, where complex communication patterns and protocols need to be verified for correctness and security.

Over recent years, DSE has advanced significantly in scalability and precision, making it effective for large-scale systems like AI and machine learning infrastructures. For instance, ParaDySE\cite{b1} has introduced automated execution heuristics that improve path selection, significantly increasing branch coverage and uncovering critical bugs. Similarly, recent work by Luo et al.\cite{b11} leveraged SMT solver time prediction to reduce the overhead associated with expensive constraints, thereby improving the symbolic execution process under strict time budgets. While these approaches optimize the symbolic execution workflow in terms of search and constraint-solving, they do not directly address the cost and efficiency of the IR used for analyzing complex network protocols and AI training workloads. Zhang et al.~\cite{b6} conducted a comprehensive study showing that conventional compiler optimizations often degrade symbolic execution performance, highlighting the need for DSE-aware IR transformations. Tu et al.~\cite{b7} addressed this gap with FastKLEE, eliminating unnecessary memory-bound checks in LLVM IR using type safety inference, reducing instruction overhead and improving path throughput.

Other research has explored optimizing DSE at the execution layer for networked environments. Poeplau and Francillon~\cite{b12} proposed SymCC, a compiler-based symbolic executor that injects symbolic logic during compile-time to bypass the inefficiencies of interpreting IRs like LLVM or VEX. This yielded major speedups, underscoring IR's importance in DSE. In distributed AI systems, where network protocols and communication patterns play a significant role.\looseness=-1 


\subsection{LLM for Network Optimization}

LLM have significantly impacted the field of intelligent program analysis and code transformation, offering novel solutions for optimizing network protocols and distributed systems. One notable example is IRIS~\cite{b9}, a static analysis framework that leverages GPT-4 to automatically infer taint propagation specifications, enhancing vulnerability detection beyond traditional rule-based methods. This capability is particularly beneficial in the context of networked systems, where the complexity of communication patterns and the need for precise data flow analysis are crucial for identifying vulnerabilities that can compromise network security.

In a similar vein, Fuzz4All~\cite{b10} employs LLMs to generate high-coverage inputs across multiple languages through autoprompted strategies, discovering a wide range of new bugs in various systems. This approach holds significant promise for enhancing the security and performance of networked applications, where fuzz testing can uncover unexpected vulnerabilities in distributed systems that rely on complex communication protocols and data exchange mechanisms. Huang et al.~\cite{b13} demonstrated that LLMs outperform traditional patch-generation tools by producing semantically accurate bug fixes across multiple languages, including those commonly used in networked environments. These advancements highlight the potential for LLMs to automate low-level transformations in network protocols, such as IR rewriting and symbolic cost reduction, which are essential for optimizing performance and security in distributed AI systems and networked applications.



\section{\textsf{LIFT} Design}

The complexity inherent in IRs such as VEX poses significant challenges to the efficiency of symbolic execution.  Traditional symbolic execution frameworks, including angr, typically focus on the basic task of translating IR without actively optimizing it. This approach can lead to suboptimal performance, especially in the context of analyzing large and complex binaries with intricate control flows and memory access patterns. To address these inefficiencies, we propose a smart, hybrid two-stage optimization framework that integrates the power of LLMs into the IR optimization process. 
We name our system \textsf{LIFT} (\textbf{L}arge-language-model \textbf{I}ntegrated \textbf{F}unctional-equivalent-IR \textbf{T}ransformation). 

 As  illustrated in \autoref{fig:ir_manual}, 
\textsf{LIFT} consists of two main phases: IR Analysis and Optimization, and Symbolic Execution and Validation. In the first phase, we symbolically execute the program to extract platform-independent IRSBs and measure their performance costs using context-aware instrumentation. The most time-intensive blocks are identified and prioritized for optimization. These blocks are transformed using an LLM-powered pipeline that simplifies statements while preserving program logic. In the second phase, execution benchmarking is performed to evaluate the effectiveness of the optimizations and ensure their long-term applicability. This involves comparing runtime, path coverage, and solver invocation statistics before and after optimization. Static metrics, such as the number of IR statements, instruction types, and temporary variables, are also collected to assess the broader impact of the optimizations. Afterward, semantic verification ensures that the transformed IR is functionally equivalent to the original by comparing components at a granular level. LLMs assist in this process, enhancing scalability and generalizability, particularly for larger binaries, though with a slight trade-off in precision compared to manual methods.

\subsection{IR Analysis and Optimization}


\vspace{2mm}
\noindent\textbf{(I) IR Extraction and Profiling.}
We begin by symbolically executing the program under test, which involves converting the binary into platform-independent IR blocks, known as IRSBs. These blocks serve as normalized representations of the program’s behavior. During symbolic execution, we apply context-aware instrumentation that dynamically measures the performance cost of each IRSB. This allows us to capture the execution time of individual IR blocks over multiple runs, ensuring that we obtain stable and reliable average measurements. Once we have gathered sufficient data, we analyze the performance costs across the IRSBs to identify the most time-intensive blocks. These identified blocks, which have the greatest impact on overall performance, are then prioritized for optimization, with modifications made based on their contextual significance within the program’s flow. This approach ensures that optimization efforts are focused on the parts of the program that will yield the greatest improvement in execution efficiency.

\vspace{2mm}
\noindent\textbf{(II) IR Optimization.}
After identifying critical blocks, we plan to modify the corresponding IR statements. Our approach involves replacing time-consuming statements with simpler, semantically equivalent ones. For instance, we can eliminate no-op assignments and merge adjacent memory operations that do not alter the program's state or affect symbolic reasoning.

While manual IR rewriting offers precision and control, it becomes less scalable for larger programs or extensive codebases. To overcome this challenge, we introduce an automated optimization pipeline powered by a LLM, which is responsible for rewriting individual IR statements. Specificially, 
each selected statement is processed individually by the language model. The prompt explicitly instructs the model to generate a functionally equivalent, shorter, or more efficient version of the IR statement while preserving its semantic meaning. To prevent errors such as hallucinated commentary or syntactic mistakes, the model's output is filtered and cleaned. 

Once the optimized versions of the statements are produced, they are placed back in their respective blocks in lieu of the original expensive statements. Every altered IR block is output into a new file. These transformed blocks preserve the program's original control structure and maintain the order of execution, enabling us to compare behavior before and after optimization.
It is important to note that these transformations are applied at the statement level, without altering control flow. This ensures that the IR is optimized while maintaining the integrity of the original program's logic. \looseness=-1

\subsection{Symbolic Execution and Validation}


\vspace{2mm}
\noindent\textbf{(I) Execution Benchmarking.}
We collect all IR statements across all basic blocks and simulate their symbolic execution cost using a model. The goal of this step is not limited to optimizing the current analysis but rather to leverage a broader insight: for semantically similar blocks, if we can identify the simplest equivalent, we can reuse that simplified version across multiple instances. This allows for greater efficiency and consistency across different analyses.

In practice, statements that involve complex arithmetic operations, memory stores, or symbolic variable assignments are assigned higher weights, reflecting their higher cost in symbolic execution. Based on this cost model, we select the top-ranking statements for transformation.
Specifically, by comparing runtime, path coverage, and solver invocation statistics, we determine if the optimization has met its intended performance improvements. Additionally, we collect static metrics, including the total number of IR statements before and after transformation, the frequency of specific instruction types, and the count of unique temporary variables. This comprehensive evaluation allows for a systematic analysis of IR complexity and its influence on execution efficiency, ensuring that the transformations not only improve runtime but also reduce the overall complexity of the symbolic execution process.

By focusing on reusing the simplest, semantically equivalent IR statements, we can optimize across a wider range of scenarios, improving both efficiency and scalability.




\vspace{2mm}
\noindent\textbf{(II) Semantic Verification.}
Once a block has been rewritten, we verify that the modified IR is functionally identical to its original form. To achieve this, both the original and transformed IR are converted into structured formats and compared at the component level. This verification process inspects variables, memory addresses, constants, and overall structure to ensure semantic consistency. If any mismatch is detected, the block is flagged for further inspection or discarded from the optimization pool.

Particularly, this step is also achieved by using LLMs. 
The use of LLMs significantly reduces human effort while enabling broader coverage across IR programs. Although this approach trades off some level of precision compared to manual rewriting, it offers strong gains in scalability and generalizability, especially for larger or unfamiliar binaries.

\begin{table*}[t]
\centering
\caption{LLM Optimization Results for 10 Binaries}
\label{tab:results}
\begin{tabular}{lcccc}
\toprule
\textbf{Program} & \textbf{Execution Time} & \textbf{IR Statement Count} & \textbf{Memory Instructions} & \textbf{Temporary Variable Count} \\
\midrule
counter       & 2.55\%  & 153 & 79  & 357 \\
branching     & 2.23\%  & 147 & 73  & 349 \\
matrix        & 1.02\%  & 143 & 71  & 12  \\
methcall      & 1.72\%  & 95  & 47  & 0   \\
objinst       & 2.13\%  & 133 & 66  & 0   \\
heapsort      & 4.52\%  & 90  & 44  & 3   \\
random        & \textbf{10.24\%} & \textbf{35}  & \textbf{15}  & \textbf{15} \\
bigtest       & \textbf{53.50\%} & \textbf{217} & \textbf{106} & \textbf{5}   \\
bigprog       & 0.75\%  & 175 & 85  & 3   \\
complexprog   & 1.77\%  & 222 & 109 & 136 \\
\bottomrule
\end{tabular}
\end{table*}

\section{Evaluation}


\subsection{Experiment Setup}

\noindent\textbf{Dataset.}
We tested 10 binaries, each highlighting performance challenges closely tied to the network communication patterns in large-scale AI systems. For instance, binaries like counter and matrix represent lightweight tasks that, while simple, still require efficient data transfers and synchronization in distributed systems to avoid network congestion. In contrast, more complex binaries such as bigtest and random simulate resource-intensive computations, underscoring the need for advanced topologies and routing protocols to handle large datasets and mitigate communication bottlenecks. Additionally, binaries like heapsort and complexprog involve dynamic memory management and multi-step data processing. These insights reflect the growing needs of modern AI infrastructure, where optimized network communication, innovative topologies, and sophisticated synchronization techniques are essential to support large-scale distributed training clusters for generative AI models.

\vspace{2mm}
\noindent\textbf{Environment.} Each binary was analyzed using a pipeline that extracted and transformed high-cost VEX IR statements via GPT-4o, followed by reintegration and evaluation within a symbolic execution engine. 



\vspace{2mm}
\noindent\textbf{Methodology.}
We extracted and parsed the VEX IR from each binary, followed by the application of a symbolic cost model to identify high-cost IR statements based on their complexity and instruction type. These statements were then optimized using ChatGPT-4o, with the generated outputs sanitized and validated for syntactic correctness. The optimized statements were reintegrated into the corresponding IR blocks, and the updated IR files were saved. Finally, symbolic execution was re-run using angr to measure the execution time and evaluate the structural changes introduced by the optimizations.

\vspace{2mm}
\noindent\textbf{Criteria.}
The following metrics were used to compare the original and optimized binaries:

\begin{itemize}
  \item \textbf{Execution Time}: Average runtime across 100 symbolic runs per binary.
  \item \textbf{IR Statement Count}: Total number of VEX statements.
  \item \textbf{Memory Instructions}: Count the number of operations.
  \item \textbf{Temporary Variable Count}: Number of unique temporaries used in the IR.
\end{itemize}

The goal of the experiment was to assess whether LLM-generated transformations lead to structural simplification and symbolic execution performance improvements while preserving correctness.

\subsection{Experiment Results}
\begin{figure}[htbp]
  \centering
  \includegraphics[width=\linewidth]{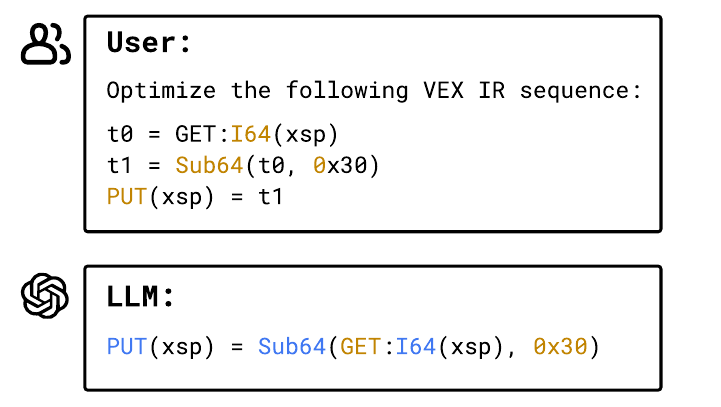}
  \Description{A diagram showing the original VEX IR and its simplified version generated by LLM, highlighting structural optimization.}
  \caption{An example of LLM optimization of VEX IR}
  \label{fig:example_llm}
\end{figure}

We observed consistent structural improvements and performance gains across all ten benchmark programs following automated IR optimization using LLMs. Figure~\ref{fig:example_llm} provides an example where a multi-line VEX IR sequence is simplified into a single statement without altering its semantics. Table~\ref{tab:results} summarizes the reductions in execution time, total IR statements, memory instructions, and temporary variables.

\noindent\textbf{Performance Improvement.}
Execution time improved in all tested programs, with the largest gain observed in bigtest, where symbolic execution time was reduced by 53.5\%. This program also saw a reduction of 217 IR statements and 106 PUT operations. Table~\ref{tab:bigtest_detail} provides a breakdown of these improvements.
Similarly, random showed a notable 10.24\% execution time reduction, along with a decrease of 35 IR statements and 15 temporary variables.

Other programs exhibited more moderate improvements. For instance, matrix and methcall experienced execution time reductions of 1.02\% and 1.72\%, respectively, while still benefiting from structural simplification. Even in the case of bigprog, where the execution time improvement was only 0.75\%, the IR was reduced by 175 statements and 85 PUT instructions.

\begin{table}[htbp]
\centering
\caption{Detailed Optimization Results for bigtest}
\label{tab:bigtest_detail}
\resizebox{\linewidth}{!}{
\begin{tabular}{lccc}
\toprule
\textbf{Metric} & \textbf{Before} & \textbf{After} & \textbf{Change} \\
\midrule
Execution Time                & 11.124952 & 5.169818 & 53.5\% ↓ \\
IR Statement Count            & 2532      & 2315     & 217 ↓ \\
PUT Instruction Count         & 542       & 436      & 106 ↓ \\
Temporary Variable Count      & 106       & 101      & 5 ↓ \\
Max Temp Variable Index       & t105      & t103     & 2 ↓ \\
\bottomrule
\end{tabular}
}
\end{table}

\noindent\textbf{Structural Simplification.}
All binaries demonstrated a reduction in IR complexity:
\begin{itemize}
    \item IR statement counts decreased by 90 to 222 in most of the programs.
    \item The number of PUT instructions dropped by up to 109.
    \item Temporary variable usage was reduced significantly in programs, like counter and branching, where over 300 temporaries were eliminated.
\end{itemize}

Semantic correctness was maintained in all optimized programs, as validated through structured comparison of IR outputs. This confirms that the LLM-generated replacements preserved the original behavior while simplifying the internal representation.

These results confirm the effectiveness of using LLMs for automated IR optimization. While the performance gains vary across program types, the transformation pipeline consistently reduced symbolic execution overhead and simplified the internal representation in all evaluated cases, all while preserving functional correctness. This demonstrates promising potential for integrating LLMs into static analysis and symbolic execution workflows.

\section{Discussion}

Our findings demonstrate that LLMs can be effectively leveraged to automate IR optimization in symbolic execution frameworks. Compared to manual optimization, the automated LLM pipeline achieves significant improvements in scalability, while still offering measurable gains in both runtime efficiency and structural simplification. The most prominent performance improvements were observed in programs with extensive symbolic arithmetic or memory access patterns. For example, the bigtest program achieved a 53.5\% reduction in symbolic execution time, along with a marked decrease in IR statement count and memory instructions. Similarly, the random program showed consistent reductions across all metrics, reinforcing the effectiveness of prioritizing high-cost statements for optimization. Even in smaller programs like methcall and matrix, IR size and temporary variable usage decreased, showing broad applicability.

A key strength of the framework lies in its modularity. By isolating individual high-cost IR statements and rewriting them independently, the system avoids the risk of introducing semantic inconsistencies across blocks. The integration of semantic validation steps ensures correctness, while the symbolic cost model enables the selective targeting of performance-critical components. However, this also introduces a limitation: optimizing statements in isolation may overlook opportunities for more holistic block-level or control-flow-sensitive transformations. Another challenge is the reliance on LLM output quality. Occasional LLM errors require post-processing, but overall output remains correct and scalable. The precision of prompt engineering and the design of validation filters play a crucial role in maintaining the reliability of the pipeline. These optimizations are well-suited for symbolic execution in distributed AI systems. These trade-offs highlight the balance between automation and control, and suggest a promising direction for integrating machine learning with formal analysis in program optimization workflows.

\section{Conclusion}
This paper introduced LIFT, a novel framework that uses LLMs to optimize IRs in DSE. By automating IR optimization, LIFT addresses the scalability issues of traditional methods, offering a scalable solution for symbolic execution in distributed AI systems. The two-phase approach: IR optimization followed by symbolic execution and validation, which ensures effective and generalizable improvements. Experimental results demonstrated significant performance gains, including reduced execution time, IR statements, and temporary variables across multiple programs. These results highlight the potential of LLMs to enhance symbolic execution and improve program analysis in complex environments.

\section*{Acknowledgements}
This study was supported by Key R\&D Program (Soft Science Project) of Shandong Province, China (2024RZB0406), the National Natural Science Foundation of China (No. 62302266, 62232010, U23A20302, U24A20244), the Shandong Science Fund for Excellent Young Scholars (No.2023HWYQ-008), the project ZR2022ZD02 supported by Shandong Provincial Natural Science Foundation and the Guangdong Basic and Applied Basic Research Foundation (Grant No. 2025A1515010111).




\bibliographystyle{ACM-Reference-Format}
\bibliography{ref}

\end{document}